\documentclass[sigconf,nonacm,screen]{acmart}

\acmConference[COTB'20]{Computer on the Beach}{01--03 de Abril de 2020}{Balneário Camboriú, SC, Brasil}

\startPage{1}

\usepackage{acronym}
\usepackage{subfloat}
\usepackage{caption}
\usepackage{subcaption}

\acrodef{IoU}{Intersection over Union}
\acrodef{GPU}{Graphics Processing Unit}
\acrodef{CNN}{Convolutional Neural Network}

\begin{document}

\title{Towards a Complete Pipeline for Segmenting Nuclei in Feulgen-Stained Images}

\author{Luiz Antonio Buschetto Macarini}
\affiliation{
  \institution{Universidade Federal de Santa Catarina -- UFSC}
  \city{Departamento de Automa{\c c}{\~a}o e Sistemas}
}
\email{luiz.buschetto@posgrad.ufsc.br}

\author{Aldo von Wangenheim}
\affiliation{
  \institution{Universidade Federal de Santa Catarina -- UFSC}
  \city{Departamento de Inform{\'a}tica e Estat{\'i}stica}
}
\email{aldo.vw@ufsc.br}

\author{Felipe Perozzo Dalto{\'e}}
\affiliation{
  \institution{Universidade Federal de Santa Catarina -- UFSC}
  \city{Departamento de Patologia}
}
\email{felipe.daltoe@ufsc.br}

\author{Alexandre Sherlley Casimiro Onofre}
\affiliation{
  \institution{Universidade Federal de Santa Catarina -- UFSC}
  \city{Departamento de An{\'a}lises Cl{\'i}nicas}
}
\email{alexandre.onofre@ufsc.br}

\author{Fabiana Botelho de Miranda Onofre}
\affiliation{
  \institution{Universidade Federal de Santa Catarina -- UFSC}
  \city{Departamento de An{\'a}lises Cl{\'i}nicas}
}
\email{fabiana.onofre@ufsc.br}

\author{Marcelo Ricardo Stemmer}
\affiliation{
  \institution{Universidade Federal de Santa Catarina -- UFSC}
  \city{Departamento de Automa{\c c}{\~a}o e Sistemas}
}
\email{marcelo.stemmer@ufsc.br}

\renewcommand{\shortauthors}{Macarini et al.}

\begin{abstract} 
Cervical cancer is the second most common cancer type in women around the world. In some countries, due to non-existent or inadequate screening, it is often detected at late stages, making standard treatment options often absent or unaffordable. It is a deadly disease that could benefit from early detection approaches. It is usually done by cytological exams which consist of visually inspecting the nuclei searching for morphological alteration. Since it is done by humans, naturally, some subjectivity is introduced. Computational methods could be used to reduce this, where the first stage of the process would be the nuclei segmentation. In this context, we present a complete pipeline for the segmentation of nuclei in Feulgen-stained images using Convolutional Neural Networks. Here we show the entire process of segmentation, since the collection of the samples, passing through pre-processing, training the network, post-processing and results evaluation. We achieved an overall IoU of 0.78, showing the affordability of the approach of nuclei segmentation on Feulgen-stained images. The code is available in: \url{https://github.com/luizbuschetto/feulgen_nuclei_segmentation}
\end{abstract}


\renewcommand{\headrulewidth}{0.5pt}
\fancyhead[L]{\textbf{XI Computer on the Beach} \\ \textit{1 a 3 de Abril de 2020, Baln. Cambori\'u, SC, Brasil}}
\fancyhead[R]{\shortauthors}
\fancyfoot[C]{\thepage}

\maketitle

\fancypagestyle{plain}{
\fancyhf{} 
\renewcommand{\headrulewidth}{0pt} 
\renewcommand{\footrulewidth}{0pt} 
\renewcommand{\footruleskip}{7mm} 
\fancyfoot[L]{\footnotesize \textbf{Formato de Refer{\^e}ncia:}
MACARINI, Luiz A. B. et al. Towards a Complete Pipeline for Segmenting Nuclei in Feulgen-Stained Images: Anais do Computer on the Beach. In COMPUTER ON THE BEACH (COTB'20), 11., 2020, Balne{\'a}rio Cambori{\'u}. \textbf{Anais...} Balne{\'a}rio Cambori{\'u}: Universidade do Vale do Itaja{\'i}, 2020, p. 1-7.}}
\thispagestyle{plain} 

\section{Introduction} \label{sec:introduction}
Cervical cancer is the second most common cancer type in women around the world. One of the causes is that it is often detected at late stages due to non-existent or inadequate screening, and the standard treatment options are often absent or unaffordable \cite{schiffman2007human}. Cervical cancer is developed in the cervical transformation zone and most of the time is caused by the infection of several types of Human Papillomavirus (HPV). This is a curable type of cancer if early detected and treated appropriately. However, detecting this disease at the pre-cancerous stage remains a challenging task \cite{song2016accurate}.

After abnormal screening results, the prevention of cervical cancer depends on the destruction or excision of the entire transformation zone epithelium, not specific precancerous lesions. This method is effective in about 80\% to 95\% of the cases \cite{dobbs2000does, arbyn2005clinical}. Cervical cancer is a deadly disease that could greatly benefit from early detection approaches \cite{weigum2010nano}. If early detected, the prognosis for the patients is excellent, improving the survival rate. The early detection of cancer can be achieved by non-invasive methods that preserve vital organ function, resulting in a better quality life for the patients \cite{seiwert2005state}.

Chromosome instability has been gaining interest because it is a central process in the development of cancer cells. This instability is indicative of the fact that the cell contains an abnormal amount of DNA, resulting in a process known as aneuploidy. This cellular abnormality has been associated with tumorigenesis. This ploidy analysis has long been a promising and economically viable solution to facilitate early cancer detection. However, this approach has not been widely adopted in clinical routine \cite{danielsen2016revisiting}. Nowadays, the biopsy is the gold standard in the detection of cancer/pre-cancer lesions, and for being an invasive method, it is recommended to be done in extreme cases only \cite{chatterjee2018augmentation}. 

This ploidy analysis usually is done by cytological exams, which consists of visually inspecting the nuclei searching for morphological alteration. This manual evaluation of microscopic images is subject to variations in perceptions and level of expertise of the cytologists, making this process prone to human errors too. Also, the slow processing time of manual analysis should be considered as another reason for using computational methods \cite{li2017gating}. To overcome this, computer-assisted analysis can be used to measure the cytological alterations and indication of cellular status, helping on early cancer diagnosis \cite{mehrotra2011efficacy, shin2010advances}.

Advances in image analysis and machine learning make available more robust algorithms for extracting information from data \cite{kandemir2015computer}. Recently, deep learning methods have achieved good performance in various computational tasks, showing to be effective for extracting the features from data in different settings \cite{lecun2015deep}. Recent efforts to use deep learning approaches in genomics and biomedical application show their flexibility for handling complex problems. Therefore, using deep learning methods to analyze cytometry data is a very promising approach \cite{li2017gating}. 



In this context, we present a complete pipeline for cell nuclei segmentation in Feulgen stained slides using image processing and deep learning methods. More specifically, we get the whole slide images, generate patches from it, use an U-Net with ResNet34 as the backbone for semantic segmentation and post-process the results with simple morphological operations. Our main goal is to provide an overview of the entire process, aiming to provide insights on early detecting possible cases of cervical cancer.

The rest of the paper is organized as follows: in Section \ref{sec:related_works} we present the related works. In Section \ref{sec:methodology} we introduce the methodology, giving details about each step of the pipeline. In Section \ref{sec:results} the results are presented, together with the discussion. Closing this work, in Section \ref{sec:final_remarks}, we present the final remarks and future works.

\section{Related Works} \label{sec:related_works}
Medical images combined with the powerful tools provided by deep learning opens up an opportunity to bring advances in computer-assisted diagnosis. Over the literature, we could find many researchers combining these tools to propose novel solutions, and some of them are presented below.

In \cite{zhang2017deeppap} the authors presented a method to classify cervical cells without prior segmentation based on deep features, using convolutional neural networks. The neural network is trained with a natural dataset and fine-tuned on a cervical cell dataset. Also on cervical cells, in \cite{zhang2017combining} the authors combined a fully convolutional network with a graph-based approach for segmentation of nuclei in cervical images. The network was used to learn the nuclei label mask and nuclei probabilistic map. It is formulated into the graph cost function in addition to the properties of the nuclei border and nuclei region. The information of the nuclei-cytoplasm position is used to modify the cost functions, where the optimal path needs to be found.

A method to segment cervical cells from Pap Smear images with the goal of early detection of cancer was proposed in \cite{song2016accurate}. Deep convolutional neural networks were used to learn diverse cell appearance features. The high-level shape information was also used to guide segmentation. In \cite{song2014deep} the authors propose a superpixel and convolutional neural network method based on cervical cancer cell segmentation. The method first segment the cytoplasm, since its contrast with the background sometimes is not good. From this, the cell nuclei are segmented and further refined. In \cite{song2016segmenting} it was proposed a method for segmenting overlapping cells in Pap smear images. The authors define the problem as a discrete labeling task for multiple cells. So coarse labeling result is used to initialize a dynamic multiple-template deformation model for further boundary refinement on each cell. Further, multiple-scale deep convolutional networks were trained to learn cell features. High-level shape information was used to help on segmenting where the cells' boundaries are noisy or lost due to touching and overlapping cells.

A fully convolutional neural network for individual nuclei segmentation was presented in \cite{cui2018deep}. The network outputs a nuclei map and a boundary map, where the post-processing step is parameter-free and was applied on the estimated nuclei map. In \cite{akram2016cell} the authors proposed a method that can be used for cell segmentation, detection, and tracking. The approach was based in two stages. In the first, a convolutional neural network outputs the cells bounding boxes along with their scores. In the second step, a second \ac{CNN} uses the proposed bounding boxes to predict segmentation masks for the cells. 

In \cite{chen2017dcan} the authors proposed a deep contour-aware network for nuclei detection and segmentation. They use a multi-level contextual feature based on a fully convolutional neural network to deal with large appearance variations. The proposed method outputs accurate probability maps and presents the contours simultaneously in a very clear way, separating clustered object instances. A novel method consisting of \ac{CNN} based structured regression model to segment cells was presented in \cite{xie2015beyond}. The method was able to handle touching cells, inhomogeneous background noises and large variations in sizes and shapes. Instead of providing a single class label, the method generates structured outputs referenced as proximity patches, where the maximum positions indicate the cell centroids.

The authors from \cite{pan2017accurate} proposed an automated cell segmentation method to segment breast cancer histopathology images. The procedure has three stages. On the first, sparse reconstruction was applied to remove the background and enhance the nuclei. Then, a \ac{CNN} was applied to segment the cell nuclei from the background. So, morphological operations were used to improve segmentation performance. The approach presented a pixel-wise segmentation accuracy better or equivalent to state-of-art algorithms. Also on automatic nuclei segmentation, in \cite{xing2015automatic} a \ac{CNN} was used to generate a probability map and from this, an iterative region merging approach was performed for shape initialization. So, a novel segmentation algorithm was presented to separate individual nuclei, combining a robust selection-based sparse shape model and a local repulsive deformable model. 

Regarding classical methods, in \cite{kumar2016unsupervised} the authors proposed an approach that first segments the cell clumps from the cervical smear image and then, detects the nuclei in each cell clump. For an accurate segmentation of nuclei in each clump, the authors also proposed a modified Otsu method with prior class probability. The contour around each nucleus was evolved until it finds the cytoplasm boundaries by the use of distance regularized level set evolution. Also dealing with overlapping cells, in \cite{lee2016segmentation} the authors proposed an automatic segmentation method for cervical cells in microscopic images. The method was divided into three steps, where in the first, the cell masses were detected by superpixel generation and triangle thresholding. Secondly, nuclei of cells are extracted by local thresholding and outlier removal. Finally, the cell cytoplasm was segmented by superpixel partitioning and refined by cell-wise contour refinement with graph cuts. The authors in \cite{ragothaman2016unsupervised} proposed a method using Gaussian Mixture Models to segment cell regions to identify cellular features such as nuclei and cytoplasm. This method was combined with the shape-based identification of nuclei to increase the accuracy of segmentation. This allows the algorithm to trace the cells and nuclei contours from the pap-smear images that contain cell clusters. This approach also deals with inconsistent staining.

\section{Methodology} \label{sec:methodology}
\begin{figure*}[]
\centering
\includegraphics[scale=0.32]{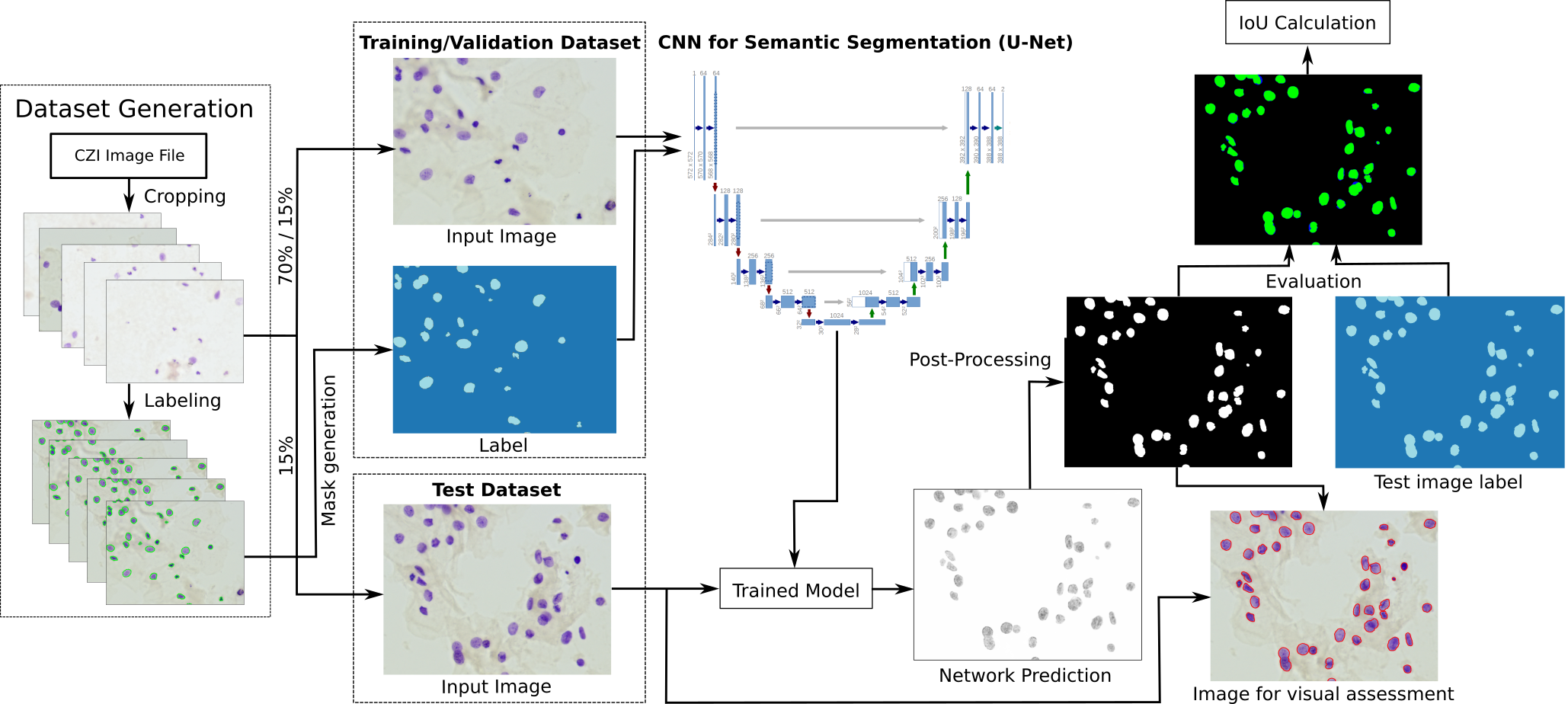}
\caption{Overview of the proposed pipeline. The U-Net Model image was adapted from \cite{ronneberger2015u}.}
\label{fig:methodology}
\end{figure*}

In this section, the methodology will be explained in detail. Fig. \ref{fig:methodology} shows an overview of the entire process. The first step consists in getting the image file of the scanned slide and submit it to the corresponding software to generate the image patches (cropping process) that will be used to train the network. Training the network for semantic segmentation is the second step, where we used the \textit{fast.ai}\footnote{https://www.fast.ai/} framework. The post-processing step consists in using image processing methods to improve the outcome of the network. The last step is the evaluation of the results, which was done in post-processed images using the \ac{IoU} metric. In the next subsections, these steps will be explained in detail. 

\subsection{Dataset Generation}

\subsubsection{Samples Collection}
The images come from exams that were performed in women who were attended at the gynecology and colposcopy outpatient clinic of the University Hospital Professor Polydoro Ernani de São Thiago of Federal University of Santa Catarina (HU-UFSC). These women, presenting cytological alterations in the oncotic cytology of previous exams, were referred to the HU-UFSC for a gynecological exam, colposcopy and biopsy\footnote{This research was approved by the UFSC Research Ethics Committee (CEPSH), protocol number 57423616.3.0000.0121. All participating patients were previously approached and informed about the study objectives. Those who agreed to participate signed an Informed Consent Form.}. The cytological samples were collected, prepared, processed and colored with Papanicolaou stain. After cytological analysis, the slides were immersed in xylol until complete removal of the coverslips for subsequent Feulgen staining. It is a widely used cytohistochemical reaction where the staining intensity is proportional to the DNA concentration. It is mostly used for DNA quantification in cell nuclei through image cytometry, aiming to do the ploidy evaluation \cite{chieco1999feulgen}. For the acquisition step, it was used a ZEISS Axio Scanner.Z1 with a Hitachi HV-F202SCL as imaging device. The exposure time was $200\mu s$, with a depth of focus of $1.22\mu s$ and a light source intensity of $246\%$.

\subsubsection{Images pre-processing}
The patches (i.e. tiles) are generated from a cropping process from the whole slide images using the \textit{Zen Software 2.6  (Blue Edition)}. This patch generation is necessary since the whole slide image resolution is too high to be used in network training (about $209000 \times 148000$ pixels). The process consists of open the CZI (Carl Zeiss Image Data File) slide file in the software and exports the data into tiles. It will generate sub-images with the resolution of $1200 \times 1600$. From now on, we will refer to these "patches" simply by "image".

\begin{figure}[H]
     \centering
     \begin{subfigure}[b]{0.4\textwidth}
         \centering
         \includegraphics[width=\textwidth]{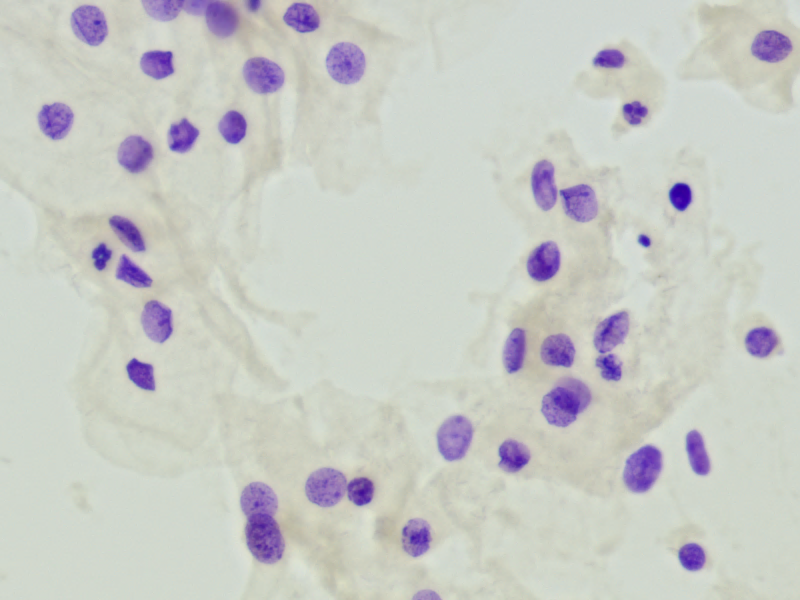}
         \caption{An input image}
         \label{fig:fig_1138}
     \end{subfigure}
     \hfill
     \begin{subfigure}[b]{0.415\textwidth}
         \centering
         \includegraphics[width=\textwidth]{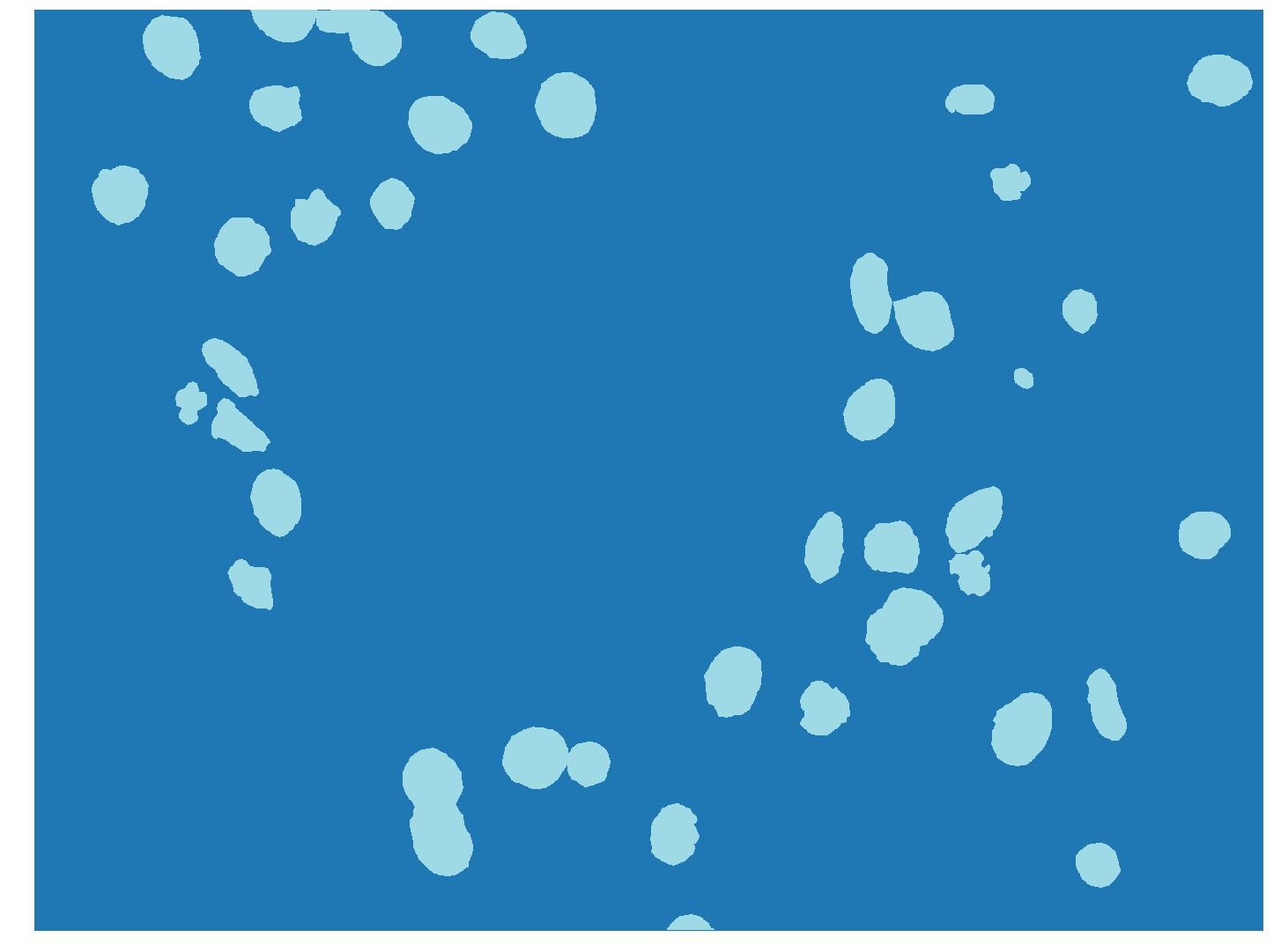}
         \caption{Mask created from specialist's labeling process}
         \label{fig:fig_1138_mask}
     \end{subfigure}
        \caption{An example of an input image and its corresponding mask}
        \label{fig:img_mask}
\end{figure}

These nuclei need to be labeled for network training. So, specialists labeled the nuclei on each image using the \textit{labelme} \cite{labelme2016} software. This software generates a JSON file with the points of the surrounding polygon that was created. The software itself provides the script to convert from JSON to Pascal VOC format. Fig. \ref{fig:img_mask} shows an image and its corresponding mask.

\subsection{Network Training}
The dataset contains 4753 images, and it was divided into 70\% from training (3327 images), 15\% (713 images) for validation and 15\% for testing. An U-Net \cite{ronneberger2015u} with ResNet34 \cite{he2016deep} as backbone was trained with a weight decay of $1e^{-03}$ and the \ac{IoU} metric as performance measure during training. We choose to use an U-Net since it was created for biomedical image segmentation purposes, presenting good results on applications over the literature.

The data augmentation applied on images was a random flip with a probability of 0.5, a random rotation between -10 and 10 degrees with probability \textit{p\_affine}, random zoom between 1 and 1.1 with probability \textit{p\_affine}, and a random symmetric warp of magnitude between -0.2 and 0.2 was applied with probability \textit{p\_affine}. There was a 0.75 probability that each lighting transform was applied and 0.2 was the probability that each affine transform was applied. Finally, a normalized transform using \textit{imagenet\_stats} was also applied.

\begin{figure}[H]
\centering
\includegraphics[scale=0.45]{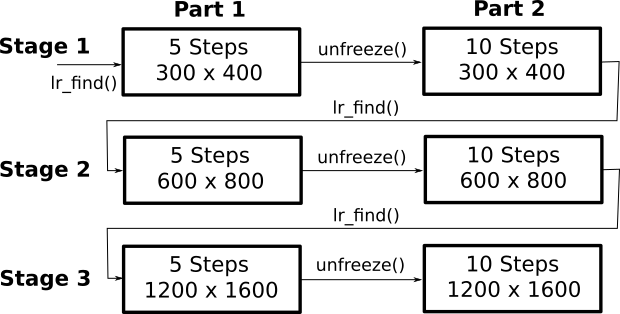}
\caption{Overview of the training process strategy}
\label{fig:training_process}
\end{figure}

The training strategy was divided into three stages, and each stage was divided into two parts, as shown in Fig. \ref{fig:training_process}. From one step to another, the image size was increased by a factor of 2. In other words, on the first stage, the network was trained with the images having 1/4 of its original size. On the second stage, training continues with the image having 1/2 of its original size and lastly, the training is completed using the images on its original size. This strategy is based on the idea that, by increasing the size of the images, we have a whole new dataset. This size increase is similar to data augmentation, which helps to avoid overfitting \cite{howard:2019}.

In the first part of each stage, the network was trained by five iterations using the \textit{fit\_one\_cycle} method, which implements the \textit{1cycle policy} \cite{abs-1803-09820}. It consists of training the model while increasing the learning rate from a very low value to a very large one, stopping when the loss starts to diverge. So, the losses are plotted against the learning rates and the chosen value is the one that is placed a bit before the minimum loss, where it can still improve. The implementation provided by fast.ai is done with the same data-augmentation as in the original paper, with a little tweak. The padded pixels in black are not colored. Instead, it is used a reflection padding \cite{gugger:2018}.

The batch size in each stage was chosen by technical limitations, where we tried several values, beginning with a large value and decreasing it by a factor of $2^n$ (32, 16, 8...) until the data fits the \ac{GPU} memory. The learning rate for each stage was found by the \textit{lr\_find} method, which tries values from $1e^{-07}$ to $10$ and stops when loss diverges. This method is provided by fast.ai and discussed in \cite{smith15a}. The learning rate value has to be chosen right before the best loss be achieved since it gives a chance to improve it during training \cite{howard:2019}.  

In the first part of each stage, only the final layers of the network were trained. So, we used the method \textit{unfreeze} to unfreeze the network and set every layer group as trainable. Again, the \textit{fit\_one\_cycle} method was used to train the network for 10 iterations, with varying learning rates, using the value found by \textit{lr\_find} method on the previous step as reference. This process will be repeated three times, where in the last one the network will be trained with the images on its original size. 

\subsection{Post-Processing}
\begin{figure}[]
\centering
\includegraphics[scale=0.25]{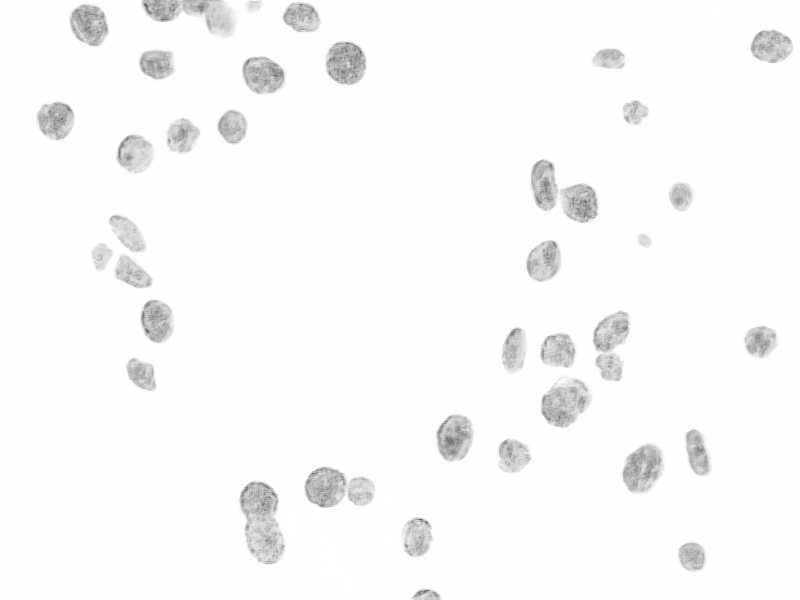}
\caption{The output of the network performed on the test set}
\label{fig:fastai_output}
\end{figure}

The trained model was used to perform the prediction on the test dataset. The fast.ai provides a result in the form of a mask, where the pixel intensity denotes the classification probability of each pixel belongs to a particular class. An example is shown in Fig. \ref{fig:fastai_output}. To improve the network's outcome, we performed a simple post-processing step to remove noise and generate another image for visualization purposes.  

\begin{figure}[H]
\centering
\includegraphics[scale=0.25]{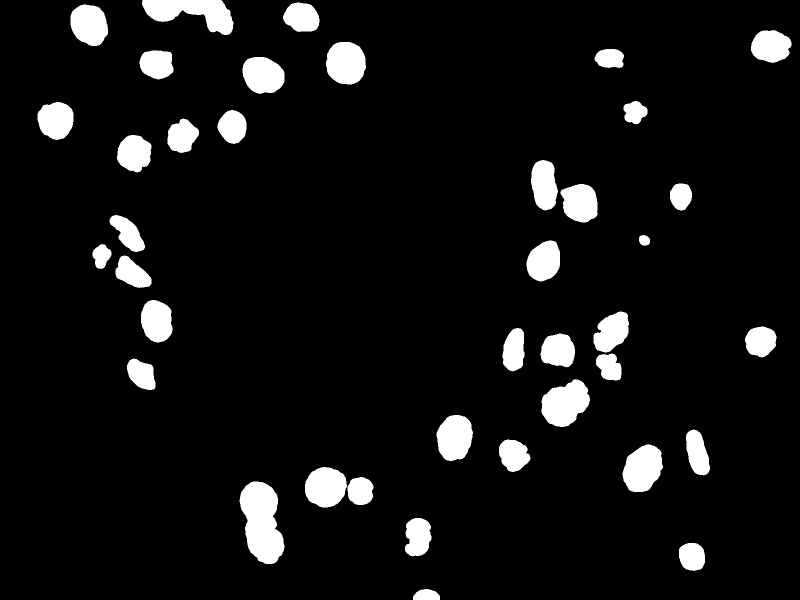}
\caption{Resulting mask of post-processing step}
\label{fig:1138_mask}
\end{figure}

We read the output mask (Fig. \ref{fig:fastai_output}) from the network and convert it to grayscale (pixel values from 0 to 255). So, a Gaussian Blur with square kernel was applied and then it was performed a threshold operation. Next, we performed an erosion operation and after, an opening (erosion followed by dilation) operation, both with a square kernel. It results in an image similar to the one shown in Fig. \ref{fig:1138_mask}. This mask is used in the network evaluation process (\ac{IoU} computation). 

\begin{figure}[H]
\centering
\includegraphics[scale=0.25]{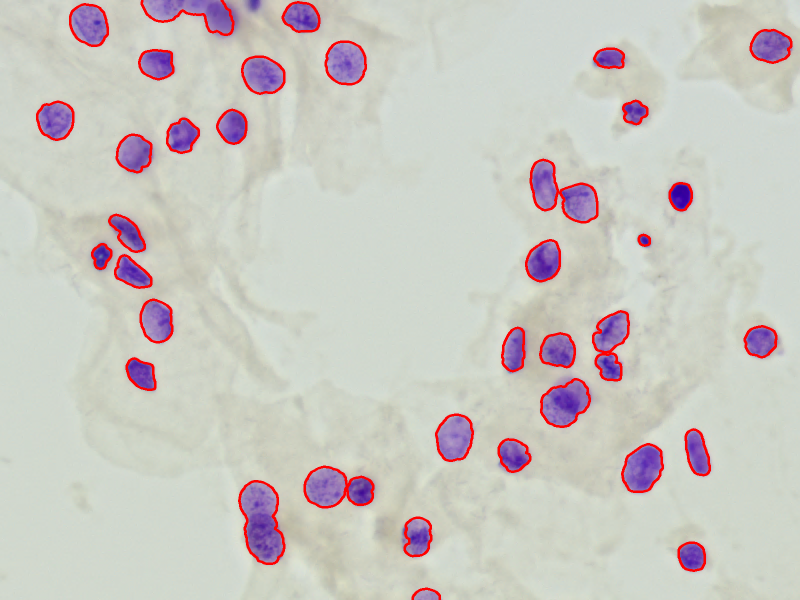}
\caption{Overlapping image created for visualization purposes}
\label{fig:1138_overlap}
\end{figure}

In order to better visualize the results, we use the Canny Operator \cite{canny1986computational} to get only the edges of the mask. These edges are overlapped on the original image, getting something similar to the image shown in Fig. \ref{fig:1138_overlap}.

\subsection{Evaluation}
We evaluated the trained model on the test dataset using the \ac{IoU} metric. The IoU metric is used to quantify the overlapping between the ground truth and the prediction output from the network. This metric is related to the Dice coefficient, a commonly used loss function during the network training. In other words, the \ac{IoU} metric is computed by the number of common pixels between the target and prediction masks divided by the total number of pixels present across both masks \cite{jordan:2018}, as shown in Equation \ref{eq:iou}.

\begin{equation}
    IoU = \frac{target \cap prediction}{target \cup prediction}
    \label{eq:iou}
\end{equation}

\begin{figure}[H]
\centering
\includegraphics[scale=0.25]{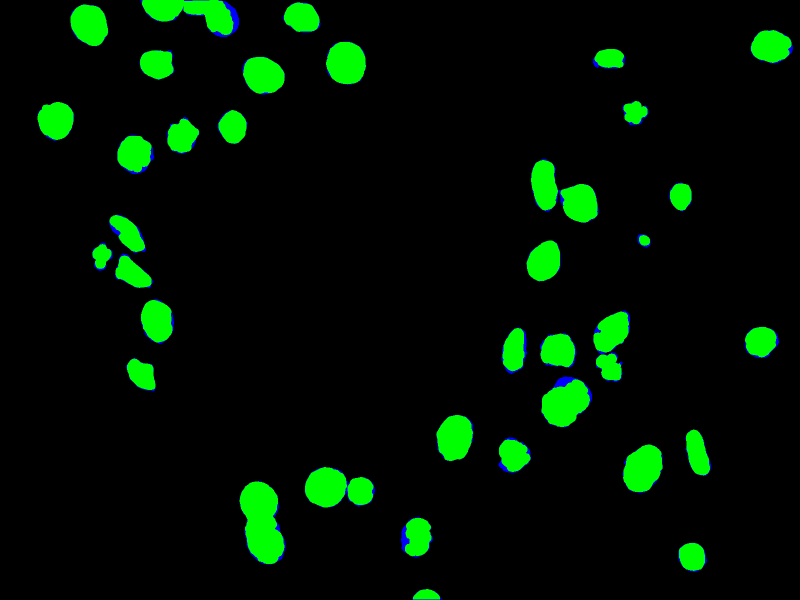}
\caption{Overlapping between the intersection (green) and the union (blue) used to calculate the IoU}
\label{fig:1138_iou}
\end{figure}

In order to illustrate this, Fig. \ref{fig:1138_iou} shows an example of IoU between the ground truth and the prediction of the network. The green pixels show the intersection between ground truth and prediction, and the blue pixels show the union between them.

\section{Results} \label{sec:results}
The network was trained on an Intel Core i7-7700 CPU @ 3.60GHz $\times$ 8, with 32 GiB of RAM, a GeForce GTX 1080 Ti using Ubuntu 18.04.3 LTS 64-bit. It was used the \textit{fast.ai 1.0.57} framework for semantic segmentation and all scripts were coded in Python 3. 

Initially, the network was trained with images having a size of $300\times400$ pixels (a quarter of its original size), batch size of 16 and learning rate of $1e^{-2}$. In the second stage of training, the images were resized to half of its original size ($600\times800$ pixels), it was used a batch size of 4 and the learning rate found was $1e^{-04}$. The last stage consists in repeating the same process with the images in its original size ($1200\times1600$ pixels). The learning rate found was $1e^{-05}$ and a batch size of 1. 
The parameters for the post-processing step were defined experimentally, with an expert doing the visual inspection of the results. It was used a $5 \times 5$ kernel for Gaussian Blur application, a $3 \times 3$ ellipse kernel for erosion operation, a $15 \times 15$ elliptical-shaped kernel for the opening operation and a threshold value of 230.

The final training loss was $0.011802$ and validation loss was $0.012477$. The \ac{IoU} was computed for each image and the average value was calculated using the entire dataset to obtain the overall \ac{IoU}. In this work, we obtained an average \ac{IoU} of $0.7841$ for the 713 test images. The entire training process took about 885 minutes (14h45m). The testing step was performed in the \ac{GPU} and took 5 minutes and 22 seconds, which gives about $0.452$ seconds per image. The post-processing step took 1 minute and 14 seconds, $0.104$ seconds per image, and was performed on CPU. 

The semantic segmentation using \ac{CNN} showed to be robust since it was capable of detecting and segmenting even some of the blurred cells. The neural network showed a good performance using only RGB images, without the application of any additional pre-processing. The fact that is possible to use an entire image as input to a \ac{CNN} is one of the main advantages of this method.

One of the biggest drawbacks of the semantic segmentation process (and deep learning in general) is the need for a big number of samples to train a model. Even though a whole slide image can be used to generate a big number of samples, it is necessary a significant number of hours to label all these nuclei. However, it is important to take into account that the \ac{CNN} works like a general-purpose feature extractor. In other words, by adjusting the network weights using backpropagation, complex filters can be generated and used to detect complex patterns \cite{krizhevsky2012}. So, developing a hand-engineered feature extractor could be a difficult process, taking into account the nature and complexity of the problem. Classical approaches found on literature showed to be less effective when compared with deep learning models since it does an automatic feature extraction \cite{kamilaris2018deep}. This characteristic fits well in this application.

Training a deep learning model usually takes a large amount of time, mostly if compared with traditional machine learning methods. However, this training step needs to be done once. Also, the testing step is fast and parallelizable, which can be performed in real-time \cite{kamilaris2018deep}. Here, even that the training time was high, the testing time was small, being about $0.452$ seconds per tile. Usually, a microscopy slide like the one used in this work takes about 40 minutes (it can vary from 40 minutes to 1h30m, depending on the number of cells) to be analyzed by a professional. Using a \ac{CNN}, this process can be done in minutes. 
The post-processing step, used to improve the outcome of the network, took about $0.10$ seconds per image. This time should decrease since this step can be parallelized on a \ac{GPU}. The main drawback of this process is the need for defining some parameters, which have to be done by experiments. However, once defined, it is not necessary to be done again.

\section{Final Remarks} \label{sec:final_remarks}
Nowadays, Deep Learning has been used to perform many tasks in image processing that was not possible some years ago. Arguably it has reached its maturity with potential and promising results, improving the state-of-the-art in object recognition, object detection, and semantic segmentation. So, we present a complete pipeline using \ac{CNN} for semantic segmentation with the goal of segmenting nuclei in Feulgen stained images. 

The obtained results validated the possibility of the usage of \ac{CNN}s to segment individual nuclei, even in some partially blurred images. This approach can be used as the first step in a nuclei classification experiment aiming to find aneuploidies. Even this approach achieved good results, it is hard to compare with other works on literature because each paper uses different pre-processing techniques, metrics, models, parameters, and datasets. Furthermore, as far as we know, there is no similar publicly available dataset that we can apply our methodology and compare the results.

The training time was quite high, but can be considered reasonably good when compared with other applications of deep learning found on literature. The necessary time to export and load the trained model, and to process the images was small. It is also necessary to point out that, in our case, the application doesn't need to be performed in real-time. 

Data plays a central role in machine learning, and even more on deep learning. So, we believe that extending the size (number of samples) of the dataset to cover most cases should help in improving the outcome of the network since deep learning methods need larger datasets. One drawback of our work is the necessity of defining parameters on the post-processing step, which influences the results. However, with a small number of experiments, good results could be achieved by simply looking at the segmentation result.

Summarizing all this, the main contribution of this work is a pipeline for nuclei segmentation using semantic segmentation techniques which work as a starting point  (not only Feulgen stained images) aiming to find aneuploidy. We also present a simple, yet effective, post-processing algorithm that improves the results, removing some small imperfections on the segmentation process. Our results also show that the training strategy proposed by \cite{howard:2019} works on this type of application, where the image resizing works as a data augmentation method. 

Future works include segment the nuclei and automatically classify it in normal or not normal. Also, testing other architectures of \ac{CNN}s, applying image processing methods in the pre-processing step and trying other post-processing ones. Some classical algorithms of segmentation can be compared with the results obtained by semantic segmentation with \ac{CNN}s. Other types of cell images (with other stains, for example) can be used to check the performance of the network. Further steps will also include the classification of segmented nuclei in normal or not-normal, aiming to early detect possible cases of cancer.

\begin{acks}
We would like to thank Coordena{\c c}{\~a}o de Aperfei{\c c}oamento de Pessoal de N{\'i}vel Superior (CAPES) for funding this work. Also, we would like to thank Dr. Adriane Pogere, for providing the samples and Ms. Ane Francyne Costa for all necessary assistance for data collection.
\end{acks}


\end{document}